\RequirePackage{fix-cm}
\documentclass[small-extended]{svjour3}       
\smartqed  

\usepackage{graphicx}
\usepackage{amsmath}
\usepackage{mathptmx}

\usepackage{enumerate}
\usepackage[misc]{ifsym}
\makeatletter
\newcommand{\Mc}{}
\DeclareRobustCommand{\Mc}{%
  M%
  \raisebox{\dimexpr\fontcharht\font`M-\height}{%
    \check@mathfonts\fontsize{\sf@size}{0}\selectfont
    {c}%
  }%
}
\makeatother
\usepackage{tabularx}
\usepackage{subcaption}
\journalname{}
\usepackage[numbers]{natbib}

\begin{document}

\title{Bedrock reconstruction from free surface data for unidirectional glacier flow with basal slip.
}

\author{Elizabeth K. \Mc George       \and
        Mathieu Sellier \and
        Miguel Moyers-Gonzalez \and
        Phillip L. Wilson
}


\institute{E. K. \Mc George  \at
              School of Mathematics and Statistics\\
              University of Canterbury\\
              Christchurch, New Zealand\\
              ORCID: 0000-0001-5707-640X 
          \and
          M. Sellier \at
          Department of Mechanical Engineering\\
          University of Canterbury\\
          Christchurch, New Zealand\\
          ORCID: 0000-0002-5060-1707
          \and
          M. Moyers-Gonzalez (\Letter) \at
          School of Mathematics and Statistics\\
          University of Canterbury\\
          Christchurch, New Zealand\\
          Tel.: +64-64-3-364-2987\\
          \email{miguel.moyersgonzalez@canterbury.ac.nz} \\
          ORCID: 0000-0003-4817-1506
          \and
          P. L. Wilson \at
          School of Mathematics and Statistics\\
          University of Canterbury\\
          Christchurch, New Zealand\\
          ORCID: 0000-0002-4563-9399
}

\date{Received: date / Accepted: date}
\maketitle
\begin{abstract}
Glacier ice flow is shaped and defined by several properties, including the bedrock elevation profile and the basal slip distribution. The effect of these two basal properties can present in similar ways in the surface. For bedrock recovery this makes distinguishing between them an interesting and complex problem. The results of this paper show that in some synthetic test cases it is indeed possible to distinguish and recover both bedrock elevation and basal slip given free surface elevation and free surface velocity.
The unidirectional shallow ice approximation is used to compute steady state surface data for a number of synthetic cases with different bedrock profiles and basal slip distributions.
A simple inversion method based on Newton's method is applied to the known surface data to return the bedrock profile and basal slip distribution.
In each synthetic test case, the inversion was successful in recovering both the bedrock elevation profile and the basal slip distribution variables.
These results imply that there is a unique bedrock profile and basal slip which give rise to a unique combination of free surface velocity and free surface elevation.

\end{abstract}

\section{Introduction}
\label{sec:intro}

Understanding cryosphere dynamics is key to modelling climate change. 
The contribution of land ice to global mean sea level (GLMS) rise for medium emissions scenarios is projected to be at least 0.10 m with some models predicting a contribution of up to 0.27 m \cite{Cazenave2013}.
Cazenave et al. (2013) identified one of the main contributors to this rise as the melting of glaciers.
Glaciers are also important socially, with millions of people in the Himalaya, Karakoram and Hindu Kush mountains relying on glacial reserves for their drinking water \cite{Rowan2018}. 
Given the potentially large impact of glacier dynamics on human livelihood, comprehensive glacier models are needed.
In particular, accurate methods for calculating the total ice mass of glaciers are required. If the bedrock profile of the glacier is known, the resulting ice thickness can be used to calculate the mass of the ice for that particular glacier. Having explicit knowledge of glacier mass can be useful and influential in policy and natural resource planning.
%
%
However, due to the difficulty of measuring the bedrock profile explicitly in many cases, it is desirable to use surface measurements and an inversion model to predict the bedrock elevation.

Surface elevation and free surface velocity data are already recorded for many ice flows and glaciers.
A number of parties collect and collate data such as the World Glacier Inventory (WGI), the United States Snow and Ice Data Center (NSIDC) and the Global Land Ice Measurements from Space (GLIMS) initiative. Data is collected in a variety of ways, primarily in-situ or via air- or space- borne craft \cite{EnvironmeantalProtectionAgency2017}.
Due to the large availability of surface data, bedrock recovery methods using these free surface measurements are particularly popular.
Another feature that can be measured or calculated from surface data is the accumulation/ablation distribution for the glacier. This distribution describes how the glacier grows/diminishes over time dues to snow/ice accumulation/ablation over time. 
This paper assumes this distribution to be measurable, though it may difficult and time consuming to do so. Field measurements can be costly \cite{Kaser2003, Hubbard2005}. Accurately predicting the accumulation rate from other measurable surface variables is an area of research in and of itself and many different methods employing a vast array of techniques have been proposed \cite[e.g.][]{Cruger2004, Lal1987, Ostrem1975, Schwikowski2013}.

Glaciers exhibit gravity-driven creep flow which is sustained by the underlying sloped geography.
Glacier ice is categorised as an incompressible, nonlinear, viscous, heat conducting fluid \cite{Greve2009} which can be described mathematically by the full Stokes flow equations together with rheological laws. 
Many methods of approximating the Stokes flow equations have been proposed in the last century. One of the most widely used approximations is the shallow-ice approximation (SIA) \cite{Hutter1981, Fowler1987}. 
In the SIA model, gravity-driven ice flow is solely balanced by basal drag neglecting longitudinal and transverse stresses, as well as vertical stress gradients \cite{Adhikari2012}.
Due to the complex nature of glacier ice flow, recovering the glacier ice thickness from only surface measurements is a non-trivial inverse problem.
Variations in recovered glacier ice thickness can be as large as the ice thickness recovered for different models. The recovered thickness is also very sensitive to input data \cite{Farinotti2017}.
In addition, inversion methods can have ill-defined solutions and may impose too many assumptions, such as the no-slip condition at the base \cite{Barcilon1993, Wilchinsky2001, Adhikari2011, Gessese2015, Heining2016}.

Imposition of a no-slip condition simplifies the inverse problem significantly and allows much faster computation.
However, basal slip is known to be influential in the flow behaviour \cite{Jiskoot2011} and so is important to include if possible.
Since the primary driving force of glacier flow is gravity, flow speed is modulated by presence, or lack thereof, of friction at the glacier-bedrock boundary \cite{Cuffey2010}.
In a temperate glacier, where high temperatures cause melt, or a thick glacier, where increased regelation causes melt, the glacier-bedrock interface is wet which can cause the ice to slide along the interface easily. 
Conversely in a glacier which has a frozen base, the ice flow is stuck to the ground and does not slide \cite{Bierman2014}.
Other factors such as till composition also impact the amount of friction at the base.
Increased velocity at the base results in a lower steady state glacier surface due to a process called dynamic thinning; the loss of ice due to accelerated ice flow into the ablation zone \cite{Bevan2015,Flament2012, Pritchard2009, Shuman2011}.
Dynamic thinning can also be caused by a steeper bedrock profile simply due to the increased contribution of gravity on the glacial flow. In addition to grounded ice flow, the basal conditions of marine ice sheets are highly important for modelling purposes. This is because basal conditions have a large affect on grounding line locations \citep{Gillet-Chaulet2020}. Uncertainties in basal conditions, can lead to models incorrectly predicting the initiation of unstable retreat.

Since the free surface of an ice flow is affected by both basal slip and bedrock topography, separating the effects of these two factors in the recovery is difficult \cite{Martin2015, Monnier2017a}.
This paper seeks to accurately recover the bedrock topography of a synthetic glacier with non-constant basal slip given known free surface elevation and velocity.
The method proposed builds on the work of \cite{Gessese2015} by modifying their method to include basal slip. This is a straight-forward method when compared to other similar research which has been conducted to date and is computationally cheap.

An example of a computationally harder method is that if \citet{Monnier2017a} who conducted a similar study regarding basal slip and bedrock topography recovery in ice sheets using the SIA. Their inverse method was relatively complex and used an elliptical linear-quadratic optimal control problem to solve for height and then computed basal slip explicitly from this. Similarly, \citet{Mosbeux2016} used an optimal control framework for ice sheets using the shallow shelf approximation. Work has also been done by \citet{Raymond2009} to estimate basal slip in ice streams using a non-linear Bayesian inverse calculation to determine the probability density for the basal properties based on the surface measurements. These techniques were applied successfully to the Rutford ice stream by \citet{Pralong2011} to recover the basal slip. In that study, a smooth basal slip was recovered for the ice stream which is physically unlikely and caused Pralong and Gudmundsson to conclude that localized variations in surface velocity are primarily caused by basal topography not basal slip. Such a conclusion is corroborated by \citet{Gudmundsson2008} who investigated ``mixing/aliasing" between basal slip and basal topography in ice stream surface measurements. While they found that the effects may be separated, small amplitude perturbations in basal slip could only be detected if they had large wavelengths in comparison to the ice height.

An overview of the governing ice flow model used is given in Sect. \ref{sec:equations}. 
Section \ref{sec:directproblem} constructs the synthetic glacier surface for a number of different synthetic cases. 
The results affirm that both basal slip distribution and bedrock profile have a significant effect on the resultant steady state surface elevation and free surface velocity. 
Section \ref{sec:inverse} gives the derivation of the recovery method proposed and the results of implementing this are given in Sect. \ref{sec:results}. 
A brief sensitivity analysis of the method to noisy surface data is given in Sect. \ref{sec:sensitivity}. 
Finally, the results are discussed in Sect. \ref{sec:discussion} and final conclusions are drawn in Sect. \ref{sec:conclusion}.


\section{Governing model}
\label{sec:equations}
This paper assumes the glacier flow dynamics are well described by the SIA.
The SIA is a simplification of the full continuity equations for a parallel sided slab on an inclined bedrock. This is done by performing a scaling analysis to obtain dimensionless field equations for the glacier flow. The small parameter used assumes the glacier extent is much larger than its thickness. Some properties of the SIA model are; (1) longitudinal and transverse stresses, as well as vertical stress gradients vanish,(2) the horizontal component of the velocity points in the direction of steepest descent of the free surface and does not change with depth, and (3) domes or troughs have no horizontal velocity. Blatter et al. \cite{Blatter2011} advise caution when applying the SIA to processes on smaller scales where the assumptions may no longer be valid, for example, anisotropic basal sliding or locally steep basal topography. 
Despite potential drawbacks, the SIA is used widely in ice flow modelling as it reduces a three dimensional flow problem into a two dimensional problem. This makes it computationally simple in comparison to higher order models such as the full Stokes equations where a full force balance has to be calculated at each step.
The SIA is typically set up with $x$-direction along the flow, the $y$-direction as the transverse direction, and the $z$-direction as the upward direction normal to the gravitational field. To simplify the testing of the inverse method proposed, this paper restricts the SIA to the unidirectional case, which omits the transverse flow.

\begin{table}
\caption{Notation.}
\label{tab:notation}
\resizebox{\linewidth}{!}{\begin{tabular}{ll}
\hline\noalign{\smallskip}
Symbol         & Meaning                                                                                  \\ \noalign{\smallskip}\hline\noalign{\smallskip}
$z$            & Vertical axis, represents height above a reference elevation                             \\
$x$            & Horizontal axis, distance along glacier from an upstream reference 
\\
$t$            & Time                                                                                     \\
$H$            & Glacier height                                                                           \\
$S$            & Glacier surface                                                                          \\
$z_b$          & Bedrock profile elevation                                                                \\
$a$            & Accumulation ablation profile for the glacier                                            \\
$u$            & Velocity profile of the glacier                                                          \\
$u_b$          & Basal velocity                                                                           \\
$u_s$          & Free surface velocity                                                                    \\
$\sigma_b$         & Basal normal stress 
\\
$\tau$       & Shear stress                                                                \\
$\tau_b$       & Basal shear stress                                                                   \\
$\beta$        & Basal slip distribution     \\ 
\noalign{\smallskip}\hline
\end{tabular}}
\end{table}

At the heart of ice flow models are the continuity equations; (1) the mass continuity equation and (2) the force balance equation. The mass continuity equation for a compressible material with density $\rho$, and velocity $v$ is given by,
\begin{align}
\frac{\partial \rho}{\partial t} + \nabla \cdot (\rho \boldsymbol{v}) = 0
\label{eq:mass_continuity}
\end{align}
and the force balance, where $\boldsymbol{g}$ is gravity and $\sigma$ the stress, is given by
\begin{align}
\nabla \cdot \boldsymbol{\sigma} + \rho \boldsymbol{g} = 0.
\label{eq:force_balance}
\end{align}

\begin{figure}
	\includegraphics[width=\linewidth]{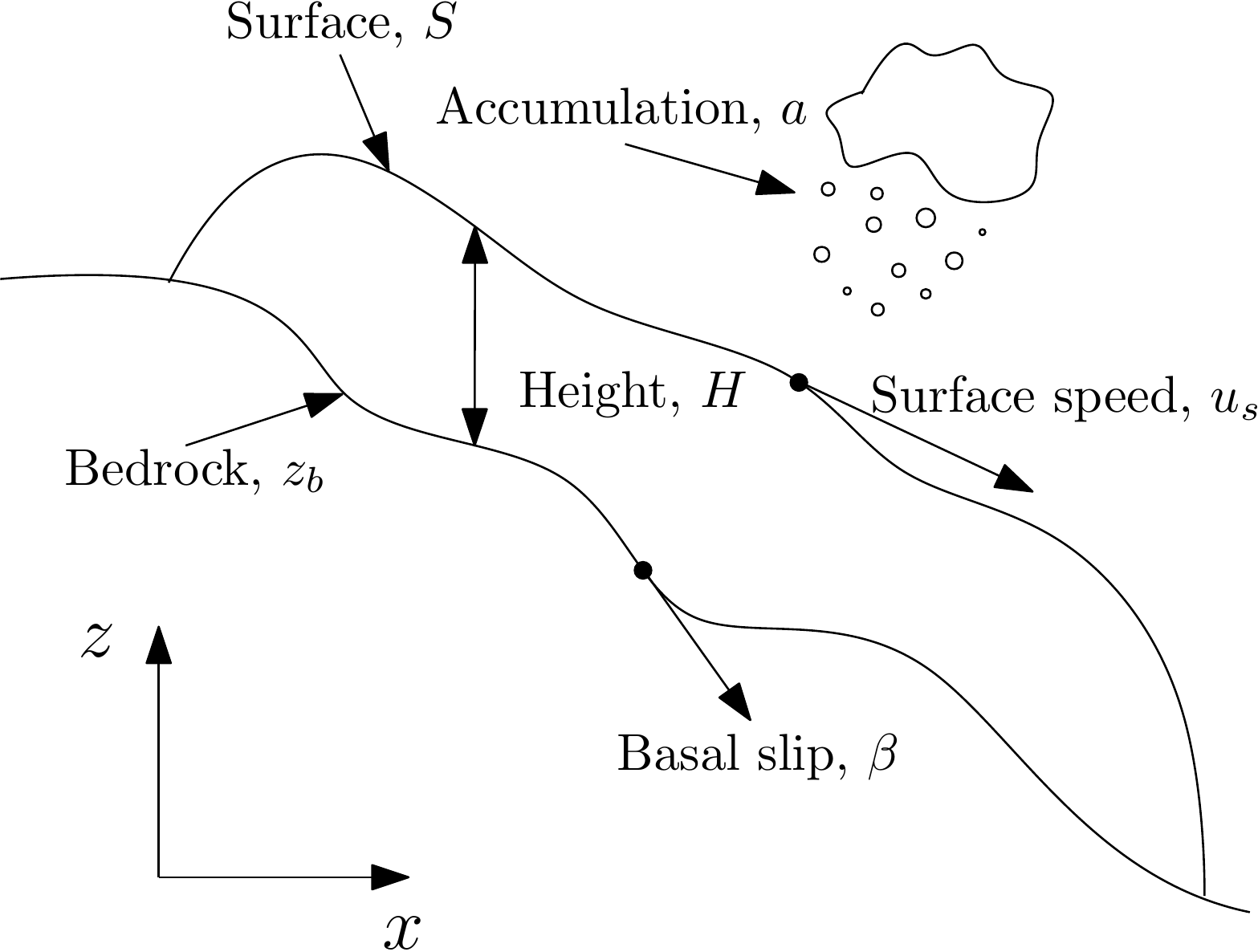}
	\caption{Glacier flowing downstream with surface $S$ and bedrock $z_b$ and thickness or height $H$. Surface speed, $u_s$ is indicated at the glacier surface and basal slip, $\beta$ is indicated at the glacier base. Accumulation, $a$, is represented as falling snow.}
	\label{fig:variables}
\end{figure}

In this paper, the ice thickness or height, $H$, is related to the surface $S$ and the bedrock elevation $z_b$ via 
\begin{align}
H = S-z_b
\label{eq:HSzb}
\end{align}
at any time, $t$. See Fig. \ref{fig:variables} for a pictorial description of this relationship.

\subsection{Stress, shear and strain in ice sheets}

Glacier flow downslope is controlled by the forces the ice experiences and the associated stresses and strains on the ice. Stress is a measure of force as applied to an area of boundary and strain is a measure of the deformation that occurs as a result of stresses.

The normal stress at the base of an ice sheet is given by
\begin{align}
\sigma_b = \rho g H
\end{align}
such that the normal stress increases linearly with the ice thickness. Calculating the component of this normal stress parallel to the slope give the basal shear stress,
\begin{align}
\tau_b = -\rho g H \frac{\partial S}{\partial x}.
\label{eq:basal_shear}
\end{align}
Glacier strain is related to the stress by Glen's empirical flow law for the shear rate \cite{Glen1952}
\begin{align}
\dot{\epsilon} &= A \tau ^n,
\label{eq:glens}
\end{align}
where $\dot{\epsilon}$ is the strain rate. As is typical for ice, we set $n=3$. The value assigned to $A$ depends strongly on temperature and should be calibrated for individual glaciers. \citet{Cuffey2010} advise caution in assuming an isotropic creep relation though admit that isotropy can be applied without compromising simulations for the overall flow in mountain glaciers. This study uses the value given in Table \ref{tab:constants} which is for an ice sheet at $-5 \deg C$ and was recommended by Cuffey and Patterson \cite[Table 3.4]{Cuffey2010}. 

\subsection{Ice velocity distribution}

In the case of a parallel sided slab, many of the strains and stresses in the ice sheet are negligible. Paired with the mass continuity equation \eqref{eq:mass_continuity} and force balance \eqref{eq:force_balance}, the result is the SIA velocity distribution given by
\begin{align}
\boldsymbol{u}(x,z) &= \frac{1}{2} A(\rho g)^3 \left (\frac{\partial S}{\partial x} \right)^3 \left[(S-z)^4 - H^4\right] + u_b(x),
\label{eq:u_x(half)}
\end{align}
where $\frac{\partial S}{\partial x}$ is the surface velocity and $u_b(x)$ the sliding velocity. Here, $\rho$ is the ice density, $g$ is the acceleration due to gravity and $A$ is the creep or flow parameter. Values for these constants are given in Table \ref{tab:constants}. The value for $\rho$ is taken as the midpoint of the range for glacier ice as recommended by Cuffey and Patterson \cite[Table 2.1]{Cuffey2010}.

The no-slip condition classically imposed \cite{Barcilon1993, Wilchinsky2001, Adhikari2011, Gessese2015, Heining2016} forces $u_b = 0$ for the glacier. 
This reduces the amount of surface data required for the inverse problem as without slip the system has only one unknown to recover. However, as discussed in the introduction, basal slip can have significant effect on glacier height
which reduces the practical applications if it is neglected.
Here, no such condition is imposed and the glacier is allowed to have varied basal slip along the base of the glacier.

Weertman \cite{Weertman1957} first proposed a power-type sliding law on a hard bed and both Fowler \cite{Fowler1987} and Lliboutry \cite{Lliboutry1968} proposed a more general form of the law for a flow with cavity formation.
Budd et al. \cite{Budd1979} found this generalised form to be empirically true for ice flow with basal sliding described by
\begin{align}
u_b = A_s \tau_b^3
\label{eq:weertman}
\end{align}
where again $\tau_b$ is the shear stress, $A_s$ the sliding constant given in Table \ref{tab:constants}, and $u_b$ the basal velocity. The value for $A_s$ is taken from Gessese et al. \cite{Gessese2015}.

Combining this sliding law \eqref{eq:weertman} with the basal shear \eqref{eq:basal_shear} gives the following expression for basal velocity
\begin{align}
u_b(x) = - \beta(x) A_s (\rho g)^3 H(x)^3 \left(\frac{\partial S(x)}{\partial x}\right)^3,
\label{eq:ub}
\end{align}
where $\beta(x)$  is the basal slip distribution which acts to regulate the amount of basal slip at the glacier base. 
Basal slip is restricted such that $\beta(x) \in [0,1]$ for all $x$ in the glacier domain. 
Physically, $\beta(x) = 0$ represents a sticky base and $\beta(x) = 1$ a friction-less base. 
It is not required for $\beta(x)$ to be constant along the glacier length. 

Hence, the full velocity distribution is given by combining (\ref{eq:u_x(half)}) and (\ref{eq:ub}),
\begin{equation}
\begin{split}
\boldsymbol{u}(x,z) = &\frac{1}{2} A(\rho g)^3 \left (\frac{\partial S}{\partial x} \right)^3 \left[(S-z)^4 - H^4\right]\\ 
&- \beta(x) A_s (\rho g)^3 H(x)^3 \left(\frac{\partial S(x)}{\partial x}\right)^3.
\end{split}
\label{eq:velocity_distribution}
\end{equation}
Note that the full velocity profile easily gives an expression for the surface velocity, $u_s(x)$, by setting $z = S$ in \eqref{eq:velocity_distribution}:
\begin{align}
u_s(x) &= - (\rho g)^3 \left (\frac{\partial S}{\partial x} \right)^3 H^3\left(\frac{1}{2} AH + \beta A_s\right).
\label{eq:us}
\end{align}
\subsection{Ice thickness}
By considering the momentum balance, volume flux, and mass conservation of the glacier, the expression for height evolution in a unidirectional ice sheet is
\begin{align}
\frac{\partial H}{\partial t} &= a(x) - \frac{\partial}{\partial x} \boldsymbol{q}(x),
\label{eq:dhdt(half)}
\end{align}
where $a(x)$ is the accumulation/ablation function of the glacier in meters of water equivalent per year, and
\begin{align}
\boldsymbol{q}(x) &= \int_{z_b}^{S} \boldsymbol{u}(x,z) dz
\label{eq:qx(half)}
\end{align}
describes the ice flux by integrating the velocity of the ice along the $x$-direction, $u_x$, from the bedrock to the free surface. Substituting \eqref{eq:velocity_distribution} into \eqref{eq:qx(half)} to gives the ice flux and finally, substituting this ice flux into the mass balance gives a non-linear diffusion equation
\begin{align}
    \frac{\partial H}{\partial t} &= a + \frac{2}{5} (\rho g)^3 \frac{\partial}{\partial x} \left(  D \frac{\partial S}{\partial x}\right)
\label{eq:dhdt}
\end{align}
with non-linear effective diffusion coefficient $D$ given by 
\begin{align}
    D = \left|\frac{\partial S}{\partial x} \right|^2 H^4 \left[ AH + \frac{5}{2}\beta A_s\right].
\end{align} 
%
\begin{table}
\caption{Typical values of constants used throughout.}
\label{tab:constants}
\begin{tabular}{lll}
\hline\noalign{\smallskip}
Symbol & Name & Value \\
\noalign{\smallskip}\hline\noalign{\smallskip}
$A_s$ & Sliding coefficient & 5 $\times 10^{-14} \text{m}^{8} \text{ N}^{-3}\text{ yr}^{-1}$ \\ 
$A$ & Glen's law parameter & 4.16 $\times 10^{-17} \text{Pa}^{-3} \text{ yr}^{-1}$\\
$\rho$ & Ice density & 880 kg m$^{-3}$ \\
$g$ & Gravitational acceleration & 9.81 m s$^{-2}$\\
\noalign{\smallskip}\hline
\end{tabular}
\end{table}

\section{Direct problem methodology}
\label{sec:directproblem}
Investigating the inverse problem requires synthetic surface data for basic ice flows.
This data is produced by solving (\ref{eq:dhdt}) for the steady state using a slight modification of the finite difference scheme as laid out by Gessese et al. \cite{Gessese2015}.
The scheme discretizes Eq. (\ref{eq:dhdt}) in time using an Euler explicit scheme and spatially using a second-order accurate central finite difference scheme. The results is
\begin{equation}
\begin{split}
	H_i^{n+1} = &H_i^n + \Delta t a_i + \frac{\Delta t}{2 \Delta x} \frac{2}{5} (\rho g)^3 \bigg\{(D_{i+1}^{n} + D_i^n) \bigg.\\ 
	&\left. \left(\frac{S_{i+1}^n - S_i^n}{\Delta x}\right) 
	 - (D_i^n + D_{i-1}^n)\left(\frac{S_i^n - S_{i-1}^n}{\Delta x}\right) \right\},
\end{split}
\label{eq:H discretisation}
\end{equation}
where
\begin{equation}
D_i = H_i^4 \left | \frac{S_{i+1} - S_{i-1}}{2\Delta x} \right | ^2 \left (AH_i + \frac{5 }{2}\beta A_s \right).
\label{eq:D discretisation}
\end{equation}
The scheme is implemented in forwards time over $\boldsymbol{x} = [0,5000]$ with a mesh size of $\Delta x = 1$ m and $\Delta t = 0.00004$ yrs. At each time step the surface is calculated via $S_i = (z_b)_i + H_i$. If, at any node, a negative height is returned, the height at that node is set to zero to reflect physical constraints. The scheme begins with no ice height across the whole domain, $H(x,0) = 0$, and then runs in forwards time until steady state is achieved as defined by
\begin{align}
    \max_i \frac{|H_i^{n+1} - H_i^n|}{\Delta t} < 0.001
\end{align}
This steady state represents an equilibrium between the accumulation and ablation due to a steady flow of ice. Note that the scheme was tested for stability with multiple other grid resolutions to confirm stability; $\Delta x = 50$ m, $\Delta t = 0.4$ yrs, $\Delta x = 50$ m, $\Delta t = 0.1$ yrs, $\Delta x = 20$ m, $\Delta t = 0.016$ yrs, and $\Delta x = 5$ m, $\Delta t = 0.001$ yrs. Each resolution produced the same steady state surface for each benchmark case as outlined in Subsect. \ref{subsec:steady_state_surfaces}. Since the method is computationally inexpensive, the finest mesh was chosen for the inverse method in Sec. 4.

\subsection{Benchmark cases}
Combinations of different accumulation/ablation rate, basal slip distribution, and bedrock elevation profile are used as benchmark cases for testing the methods.
\subsubsection{Choice of accumulation/ablation rate, $a(x)$}
For each benchmark case the accumulation/ablation function is defined as
\begin{equation}
    a(x) = \begin{cases} a_0 \left (1- \frac{300 - x}{100} \right) \quad \text{if} \quad x \leq 300 \\ a_0 \left (\frac{2200 - x}{1900} \right) \quad \text{if} \quad x \geq 300 \end{cases}
\end{equation}
where $a_0$ is the maximum value of the accumulation/ablation function and set to 0.5 for all future calculations. Adjusting this maximum values simply raises or lowers the steady state surface \cite{LeMeur2004}.
This function gives the most accumulation at the top end of the glacier and then linearly decreases along it's length until at the bottom end which has net ablation.

\subsubsection{Cases of basal slip distribution, $\beta(x)$}
\begin{figure}[ht]
	\centering
	\includegraphics[width=\linewidth]{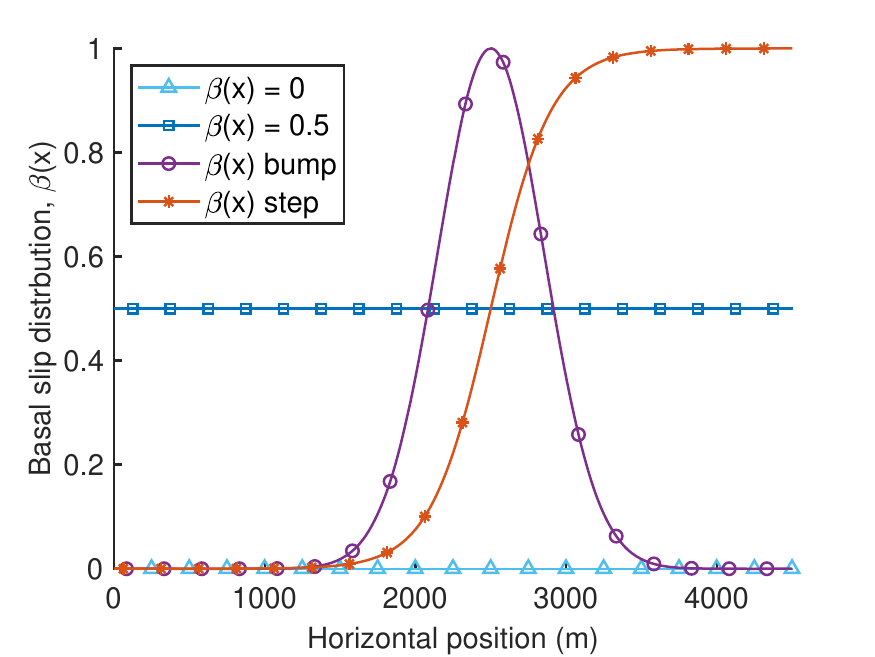}
	\caption{Examples of the three forms of basal slip distribution, $\beta(x)$. For $\beta(x)$ bump, $\gamma = 1, M = 2500, \delta = 500$. For $\beta(x)$ step, $L = 0, R = 1, M = 2500, k = 0.005$.}
	\label{fig:forms_of_beta}
\end{figure}

Three different types of basal slip distributions are tested.
The different distributions again help to test the robustness of the method against more realistic scenarios.
The three type are labelled and mathematically defined as follows. Examples of their shapes are given in Fig. \ref{fig:forms_of_beta}.
\begin{enumerate}
    \item $\beta(x)$ constant.
    
    This type of slipping gives constant slip along the glacier as described by
        \begin{align}
            \beta(x) = K,
        \end{align}
        where $K$ is a constant in $[0, 1]$. $K=0$, for example, could represent a cold-based glacier in which the ice is frozen to the bed which inhibits the flow mechanisms of sediment deformation, ice deformation and basal sliding \cite{Kleman2007, Alley1986, Iverson1995, Christoffersen2003}.
    \item  $\beta(x)$ bump.
    
    This type of slip gives negligible slip at the top and bottom of the glacier with narrow region with lots of slip in the middle as described by
        \begin{align}
            \beta(x) = \gamma  e^{\frac{ -(x-M)^2}{\delta^2}},
            \label{eq:beta_bump}
        \end{align}
        where $M$ is the midpoint of the bump, $\gamma$ is the height of the bump and $\delta$ is the extent. 
    \item $\beta(x)$ step.
    
    This type of slipping gives a transition from negligible basal slip to some slip as described by 
        \begin{align}
            \beta(x) = \frac{R}{1 + e^{k(x - M)}},
            \label{eq:beta_step}
        \end{align}
        where $\lim_{x \to \infty} \beta(x) = R$ and $\lim_{x \to -\infty} \beta(x) = 0$, $k$ is the steepness and $M$ is the midpoint of the transition between asymptotes. 
\end{enumerate}

\subsubsection{Cases of bedrock elevation profile, $z_b(x)$}
\label{subsec:cases_of_bedrock}
Two different bedrock elevation profiles are defined in the benchmark cases.
The two forms are given by:
\begin{enumerate}
    \item An inclined flat bed, $z_b = f$
        \begin{align}
            z_b(x) = f(x) = z_0 - \alpha x
        \end{align}
        where $z_0$ is the elevation at the point $x = 0$ and $\alpha$ is the slope.
    \item An inclined bumpy bed, $z_b = b$
        \begin{align}
        z_b(x) = b(x) = z_0 - \alpha x + A \sin( \lambda x);
        \end{align}
        where $z_0$ and $\alpha$ are as before, $A$ is the amplitude of the bumps and $\lambda$ wavelength. For all simulations, $A = 50$ and $\lambda = \frac{1}{350}$. This bedrock profile is designed to combine two test cases as used by Gessese et al. \cite{Gessese2015} to test their no-slip recovery method.
\end{enumerate}
In each case $z_0 = 900$ m and $\alpha = 0.2$.

\subsection{Resultant steady state surfaces for each type of basal slip.}
\label{subsec:steady_state_surfaces}
\begin{figure}[ht]
 \includegraphics[width=\linewidth]{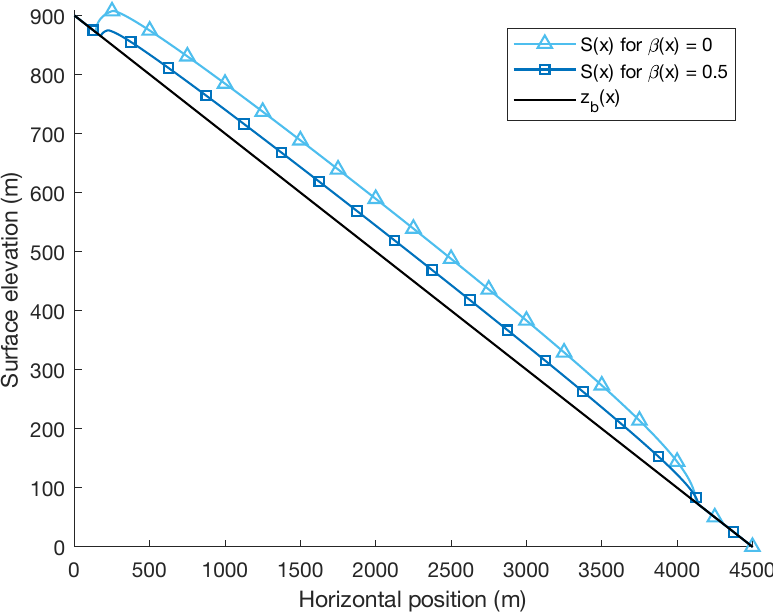}
\caption{Computed steady state glacier surfaces for constant basal slip with flat inclined bedrock, $z_b = f$.}
\label{fig:direct_const_flat}
\end{figure}

\begin{figure}[ht]
 \includegraphics[width=\linewidth]{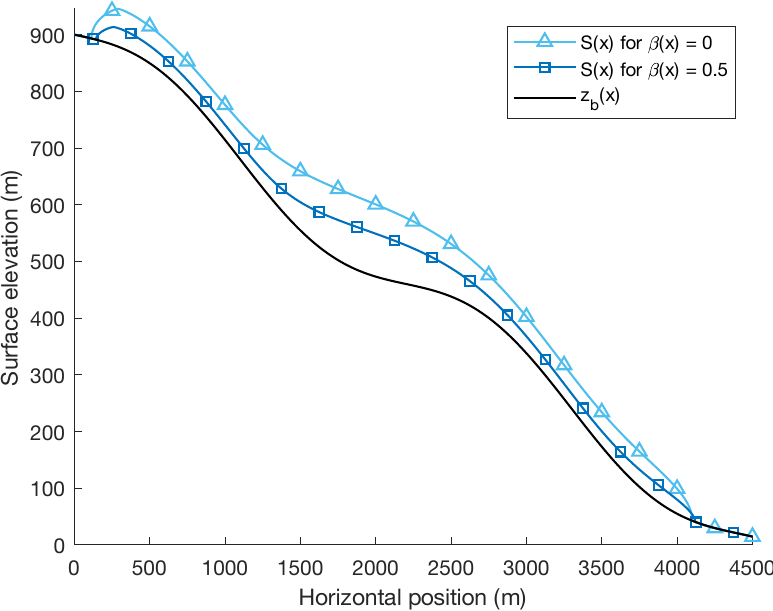}
\caption{Steady state glacier surfaces for constant basal slip with bumpy inclined bedrock, $z_b = b$.}
\label{fig:direct_const_wave}
\end{figure}

\begin{figure}[ht]
\includegraphics[width=\linewidth]{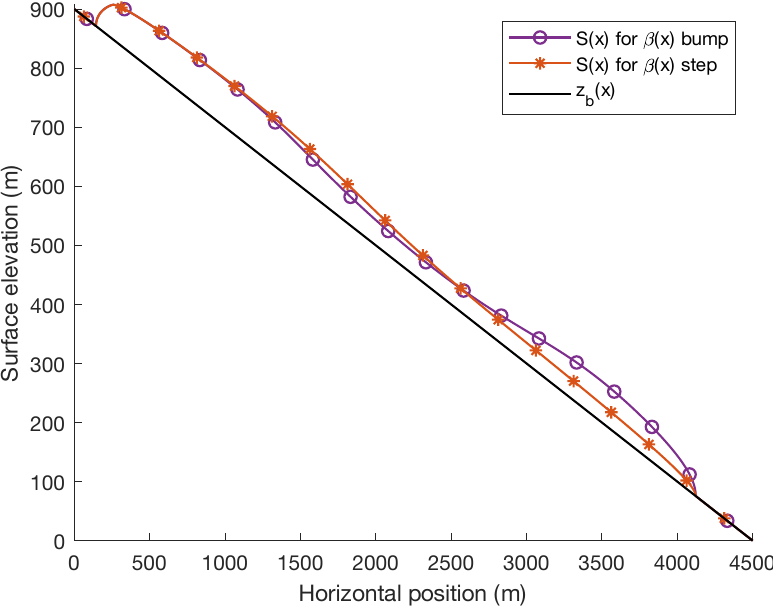}
\caption{Steady state glacier surfaces for basal slip with a bump and basal slip with a step where $z_b = f$.}
\label{fig:direct_nc_flat}
\end{figure}

\begin{figure}[ht]
\includegraphics[width=\linewidth]{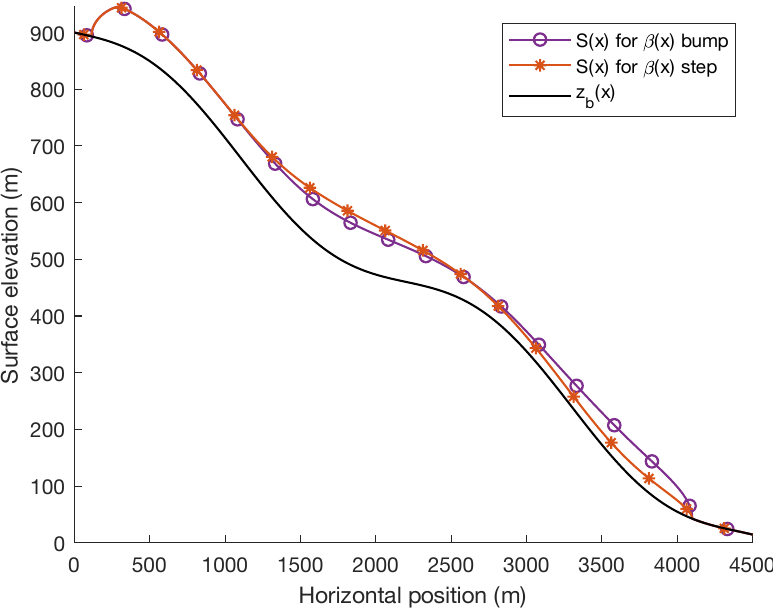}
\caption{Steady state glacier surfaces for basal slip with a bump and basal slip with a step where $z_b = b$.}
\label{fig:direct_nc_wave}
\end{figure}

Figures \ref{fig:direct_const_flat} to \ref{fig:direct_nc_wave} give the resultant steady state surfaces for each type of basal slip for the two cases of bedrock profile. In each plot, the underlying basal slip is indicated by the marker type which matches those used in Fig. \ref{fig:forms_of_beta}. 

In Figs. \ref{fig:direct_const_flat} and \ref{fig:direct_const_wave}, the case of no-slip, $\beta(x) = 0$, is plotted along side the constant slip case and is consistent with the results of Gessese et al. \cite{Gessese2015}. These two figures illustrate that increased basal slip results in a glacier with less height. This effect is consistent with literature \cite{Pritchard2009, Bevan2015} and occurs for the other types of basal slip also. Figure \ref{fig:direct_const_wave} shows that the glacier surface profile follows that of the bedrock.

In Fig. \ref{fig:direct_nc_flat}, a visible dip occurs in the glacier surface at the location of increased basal slip for the bump case. Similarly, the glacier surface is observed to lower once the transition occurs to more basal slip in the step case. In Fig. \ref{fig:direct_nc_wave} the same properties are exhibited due to changes in basal slip but are much harder to see due to the surface undulations produced by the bedrock. 

These modelled surfaces show that the shape of the glacier is affected by both the bedrock and the basal slip. Without accounting for basal slip, the dips observed in the surface will appear to be the result of related dips in the bedrock. These results for the direct case reinforce the importance of including basal slip in the recovery method for bedrock elevation.

\section{Inverse methodology}
\label{sec:inverse}
The inverse problem seeks to recover the bedrock elevation profile, $z_b(x)$ and the basal slip distribution, $\beta(x)$, for a steady state glacier from two known free surface quantities; (1) the surface elevation, $S$, and (2) the free surface velocity, $u_s$.

Glaciers and ice flows have been studied extensively throughout the past century by a variety of different groups such as NSIDC.
Data collection has been pushed particularly due to ice melt acting as a major contributor to sea level rise \citet[Fig. 13.10, 13.13]{Church2013}.
As such the body of data for glaciers and ice flows is ever increasing.
It is reasonable to assume there is, or can be measured, sufficient data for the two free surface variables as well as the accumulation ablation function.

Previous authors \cite{Gessese2015, Heining2016} have recovered bedrock data from one free surface input with the assumption of a sticky, no-slip base where $\beta(x) = 0$.  \citet{Monnier2017a} has recovered bedrock data allowing for a slipping base using a complex optimal control framework. Work has also been done to recover bedrock data for ice streams using a variety of techniques \citep{Raymond2009, Pralong2011, Gudmundsson2008}.

The inverse method proposed here is for grounded ice and seeks to compliment the simple techniques used by \citet{Gessesse2014} in their no-slip case. Where they used one surface data input, we use two input variables which allows for the recovery of bedrock elevation profile and basal slip distribution simultaneously.
As with the direct problem, the flow is assumed well described by the unidirectional SIA defined in Eq. (\ref{eq:dhdt}) and the surface velocity approximated by Eq. (\ref{eq:us}).

\subsection{Method for the inverse problem}
Given two observable variables, $u_s$ and $S$, the following will show that it is possible to accurately recover to unknown variables, $H(x)$ and $\beta(x)$. Recovery of $H(x)$ immediately gives the desired bedrock due to Eq. (\ref{eq:HSzb}). To solve for these two variables, first consider the equations which define them. Assuming a steady state surface, we set $\partial H / \partial t$ to 0  in Eq. (\ref{eq:dhdt}) giving,
\begin{align}
    0 &= a + \frac{2}{5} (\rho g)^3 \frac{\partial}{\partial x} \left(  \left|\frac{\partial S}{\partial x} \right|^2 H^4 \left[ AH + \frac{5}{2}\beta A_s\right] \frac{\partial S}{\partial x}\right)
\end{align}
which has two unknowns $H$ and $\beta$. Since the free surface velocity is also given, rearranging the equation for $\beta$ gives
\begin{align}
    \beta = \frac{-1}{A_s} \left(\frac{u_s}{(\rho g )^3 \left|\frac{\partial S}{\partial x} \right|^2 \frac{\partial S}{\partial x} H^3} + \frac{1}{2} A H \right)
    \label{eq:beta_rearrange}
\end{align}
Substituting this expression for $\beta$ into the steady state surface equation above and integrating results in
\begin{align}
0 = \int_{0}^{x} a dx - \frac{1}{10}(\rho g)^3 \left|\frac{\partial S}{\partial x} \right|^2 \frac{\partial S}{\partial x} A H^5 - u_s H + C_0
\label{eq:H_poly}
\end{align}
which is a polynomial equation in only $H$ with constant of integration
\begin{align}
    C_0 = - \int_{0}^{x_0} a dx + \frac{1}{10}(\rho g)^3 \left.\left[ \left|\frac{\partial S}{\partial x} \right|^2 \frac{\partial S}{\partial x} \right] \right|_{x = x_0} A H_0^5 +(u_s)_0 H_0
\end{align}
where $x_0$ is some point inside the domain of the glacier where the height is known. It is reasonable to assume height can be known at one location from practical measurements.

There are numerous methods which could be employed to solve the polynomial for $H$.
Newton's method is chosen for it's simplicity and controllability.
Hence solving (\ref{eq:H_poly}) for each $H_i = H(x_i)$ using Newton's method
\begin{align}
    H_i^{n+1} = H_i^{n} - \frac{F(H_i^n)}{F'(H_i^n)}
\end{align}
with the following functions
\begin{align}
F(H_i) &= \int_{0}^{x_i} a dx - \frac{1}{10}(\rho g)^3 \left.\left(\frac{\partial S}{\partial x}\right)^3\right|_{x = x_i} A H_i^5 - (u_s)_i H_i + C_0
\label{eq:f(h)}\\
F'(H_i) &= -\frac{5}{10} (\rho g)^3 \left.\left(\frac{\partial S}{\partial x}\right)^3\right|_{x = x_i} A H_i^4 - (u_s)_i .
\end{align}
For Newton's method to find the correct root of $F$ it is important to start with a nearby guess. Therefore, the method will move away from $x_0$ to the left and right using
\begin{align}
    H_i &= H_0  \quad &&\text{for } x_i = x_0 \\
    H_i^0 &= H_{i-1} \quad &&\text{for } x_i > x_0 \\
    H_i^0 &= H_{i+1} \quad &&\text{for } x_i < x_0 
\end{align}
For each glacier, $x_0$ is chosen to be in the middle of the glacier domain.

\subsection{Method for the inverse problem (non-steady state)}
\label{subsec:non-steady-state}
The inverse method above can be extended to glaciers which are not at steady state if the free surface is known at two different times, $S_1$ and $S_2$. This extra information allows the estimation
\begin{align}
\frac{\partial H}{\partial t} &= \frac{\partial S}{\partial t} \approx \frac{S_2 - S_1}{\Delta t},
\end{align}
and the modified accumulation/ablation function
\begin{align}
a' = a - \frac{\Delta S}{\Delta t}.
\end{align}
This kind of approximation technique was used by \citet{Gessesse2014} and will work in the same way here but is not investigated further.
\section{Results}
\label{sec:results}
\begin{figure*}[ht]
\centering
\begin{minipage}{0.49\textwidth}
    \centering
    \includegraphics[width=\textwidth]{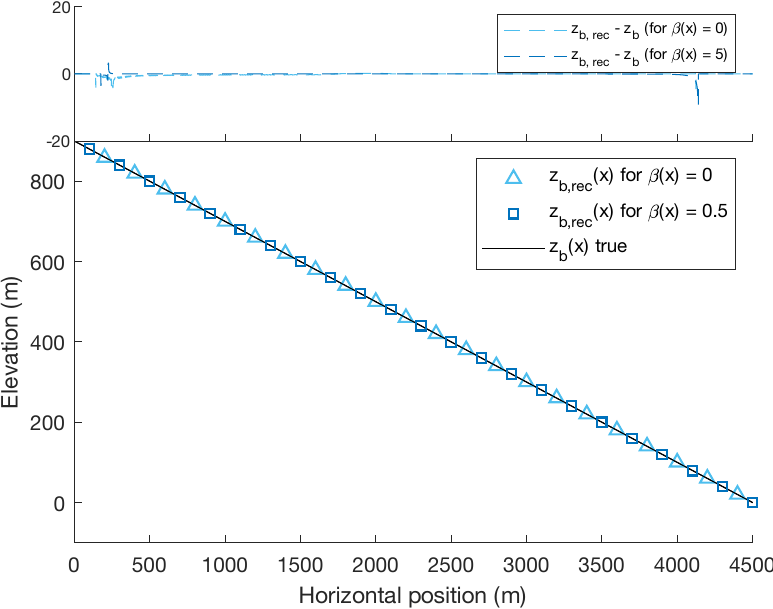}
    \subcaption[(a)]{}
\end{minipage}
\begin{minipage}{0.49\textwidth}
    \centering
    \includegraphics[width=\textwidth]{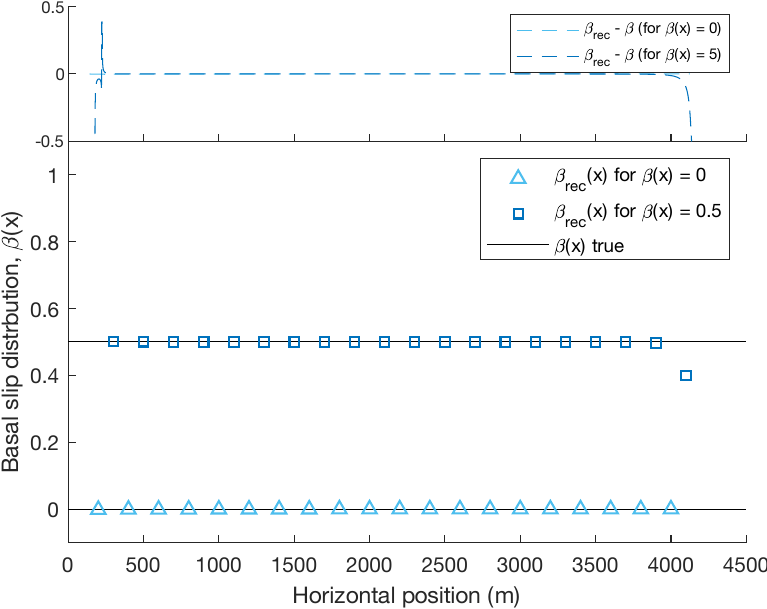}
    \subcaption[(b)]{}
\end{minipage}
\caption{Recovered bedrock (a) for non-constant basal slip where $z_b = f$ and corresponding recovered basal slip (b). 
}
\label{fig:inv_c_flat}
\end{figure*}

\begin{figure*}[ht]
\centering
\begin{minipage}{0.49\textwidth}
    \centering
    \includegraphics[width=\textwidth]{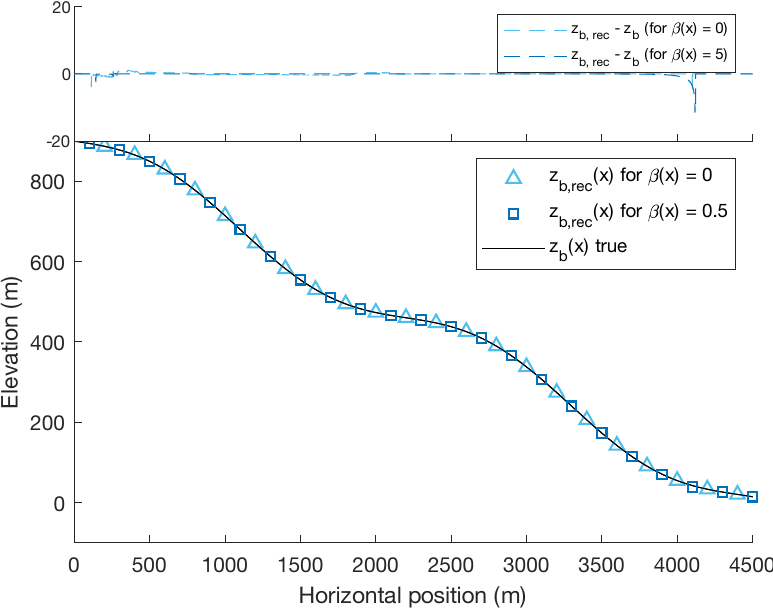}
    \subcaption[(a)]{}
\end{minipage}
\begin{minipage}{0.49\textwidth}
    \centering
    \includegraphics[width=\textwidth]{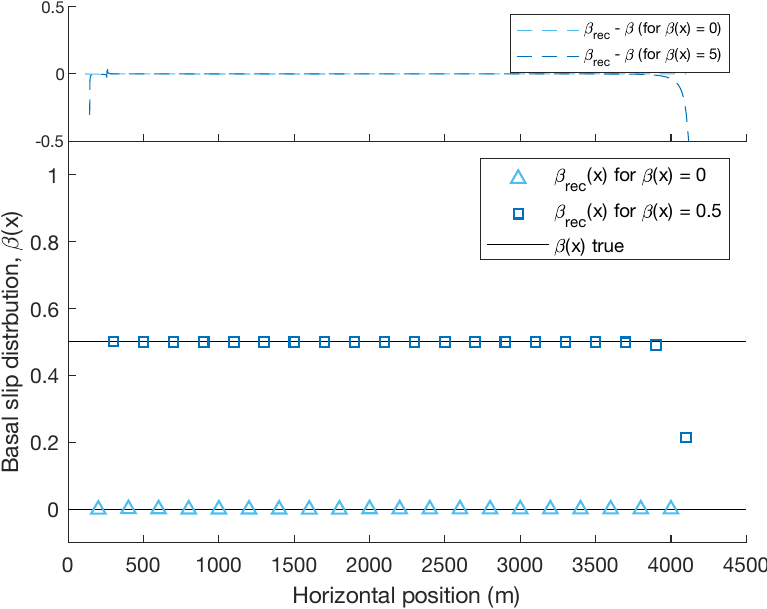}
    \subcaption[(b)]{}
\end{minipage}
\caption{Recovered bedrock (a) for constant basal slip where $z_b = f$ and corresponding recovered basal slip (b). 
}
\label{fig:inv_c_wave}
\end{figure*}

Figures \ref{fig:inv_c_flat} to \ref{fig:inv_nc_wave} show the inversion results for each combination of underlying bedrock and basal slip distribution. 
The recovered bedrock profile elevation, $z_{b, rec}$, is shown in each subfig. (a). 
The recovered basal slip distribution, $\beta_{rec}$, is shown in in each subfig. (b).
The recovered variables in each case are overlaid on the true variables for easy comparison.

Table \ref{tab:errors} gives relative errors for the recovered bedrock and basal slip distribution respectively. 
The relative error between the recovered variables, $x_{rec}$, and the true value, $x$, is calculated by
\begin{align}
    E(x_{rec}, x) = \frac{||x_{rec} - x||_2}{||x||_2} = \frac{\sqrt{\sum_i((x_{rec})_i - x_i)^2}}{\sqrt{\sum_i x_i^2}},
    \label{eq:relative_error}
\end{align}
where $i$ runs along the glacier domain. Note that (\ref{eq:relative_error}) is not defined for $x = 0$. In this case, the error is defined as
\begin{align}
    E(x_{rec}, x) = ||x_{rec}||_2  = \sqrt{\sum_i((x_{rec})_i)^2}.
    \label{eq:relative_error_0}
\end{align}

For each benchmark case, the bedrock reconstruction is in very good agreement with the true profile. The relative errors in bedrock recovery, as given in table \ref{tab:errors}, are of magnitude $10^{-3}$ or smaller for each case. This indicates that the has high accuracy for recovering bedrock elevation profiles. This is illustrated in each sub-fig. (a) which show close alignment between the recovered bedrock and the true bedrock.

For each benchmark case, the recovered basal slip distribution is in agreement with its true distribution. The relative errors, as given in table \ref{tab:errors} are of magnitude $10^{-1}$  or smaller for each case. While this is a much larger error than for bedrock recovery, it is still of small magnitude. In the (b) sub-figs., the recovered variable closely aligns with the true values. It is visible that areas or largest error are at the top and tail ends of the glacier which is to be expected.

\begin{figure*}[ht]
\centering
\begin{minipage}{0.49\textwidth}
    \centering
    \includegraphics[width=\textwidth]{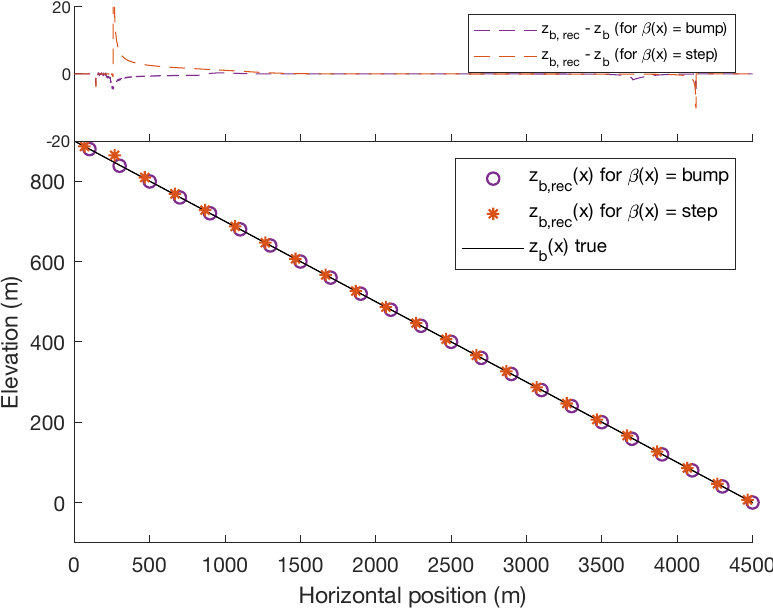}
    \subcaption[(a)]{}
\end{minipage}
\begin{minipage}{0.49\textwidth}
    \centering
    \includegraphics[width=\textwidth]{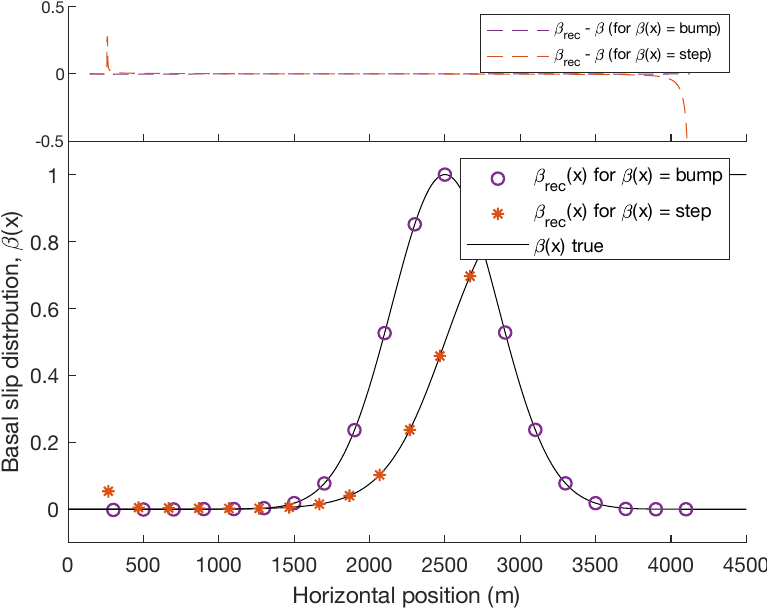}
    \subcaption[(b)]{}
\end{minipage}
\caption{Recovered bedrock (a) for non-constant basal slip where $z_b = f$ and corresponding recovered basal slip (b). 
}
\end{figure*}

\begin{figure*}[ht]
\centering
\begin{minipage}{0.49\textwidth}
    \centering
    \includegraphics[width=\textwidth]{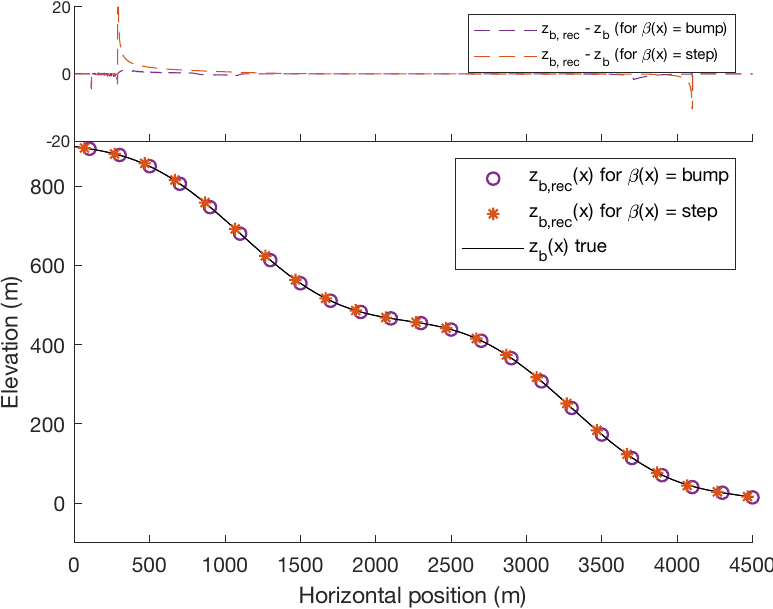}
    \subcaption[(a)]{}
\end{minipage}
\begin{minipage}{0.49\textwidth}
    \centering
    \includegraphics[width=\textwidth]{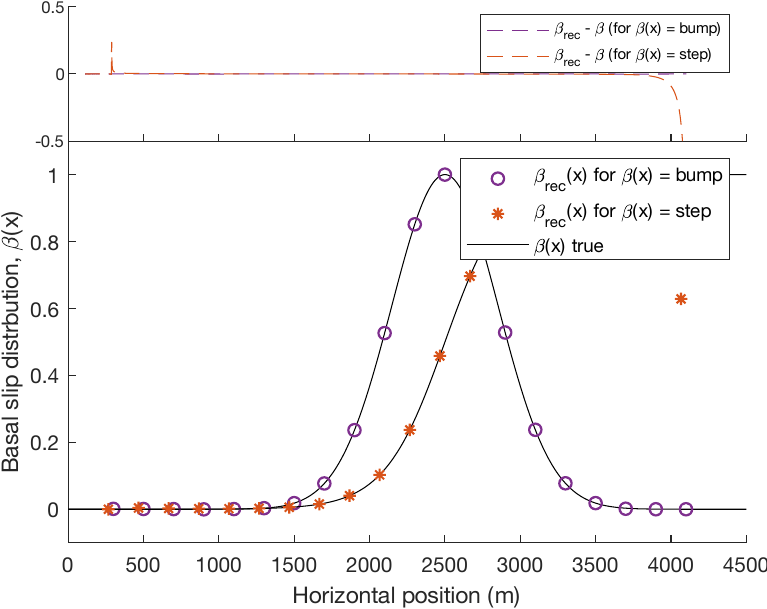}
    \subcaption[(b)]{}
\end{minipage}
\caption{Recovered bedrock (a) for non-constant basal slip where $z_b = b$ and corresponding recovered basal slip (b). 
}
\label{fig:inv_nc_wave}
\end{figure*}

\begin{table}[ht]
\caption{Associated errors for recovered bedrock profile elevation and basal slip distribution as defined by Eqs. \eqref{eq:relative_error} and \eqref{eq:relative_error_0}.}
\label{tab:errors}  
\begin{tabular}{ll|ll}
\hline\noalign{\smallskip}
Case       &       & Relative errors     &                         \\ \noalign{\smallskip}\hline\noalign{\smallskip}
$\beta(x)$ & $z_b$ & $E(z_{b, rec}, z_b)$ & $E(\beta_{rec}, \beta)$ \\ \noalign{\smallskip}\hline\noalign{\smallskip}
0          & $f$     & $7.583\times 10^{-4}$              &  $3.000 \times 10^{-2}$                \\
0          & $b$     & $5.423\times 10^{-4} $              & $2.128 \times 10^{-2}$                  \\
0.5          & $f$     & $7.733\times 10^{-4}$              & $7.192 \times 10^{-2}$                \\
0.5          & $b$     & $1.073\times 10^{-4}$              & $7.591 \times 10^{-2}$                  \\
bump       & $f$     & $9.882 \times 10^{-4}$             & $1.847 \times 10^{-3}$                 \\
bump       & $b$     & $6.509 \times 10^{-4}$             & $1.219  \times 10^{-3}$                 \\
step       & $f$     & $4.278\times 10^{-3}$             & $1.043 \times 10^{-1}    $            \\
step       & $b$     & $3.035 \times 10^{-3}$              & $1.211 \times 10^{-1}$                 \\ \noalign{\smallskip}\hline
\end{tabular}
\end{table}

\section{Sensitivity analysis}
\label{sec:sensitivity}
Surface data in ice flows in reality always has some noise and or uncertainty. Therefore, for practical applications the method should respond in predictable ways to noise.
To evaluate the effect of noise on the inversion method, noise is added to each of the measured variables, $S$ and $u_s$, as well as the accumulation/ablation function, $a$. Noise is added to synthetic data in the following way:
\begin{align}
     S_{noise} &= S_{true} + \epsilon n \max(H_{true}) \text{ and,}\\
    u_{s, noise} &= u_{s, true} + \epsilon n (\max(u_{s, true} - \min u_{s, true}),\\
    a_{noise} &= a_{true} + a_{offset} + \epsilon n (\max(a_{true} - \min a_{true}),
\end{align}
where $n \in [-1,1]$ is randomly distributed.
The amount of noise to be easily adjusted by choosing $\epsilon \in [0,1]$ where the larger the choice, the more noise. Once noise is added to surface data, the result is smoothed with a local regression using weighted linear least squares and a second degree polynomial model which assigns less weight to outliers in the regression. The local span for the regression is 20 \% of the data points. Data outside six mean absolute deviations is given zero weight.
Smoothing the data is important as $\frac{\partial S}{\partial x}$ appears regularly in the equations and needs be well defined.
Examples of noise added to surface data as well as their smoothed counterparts can be found in the supplementary material.

%

\begin{figure*}[ht]
\centering
\begin{minipage}{0.49\textwidth}
    \centering
    \includegraphics[width=\textwidth]{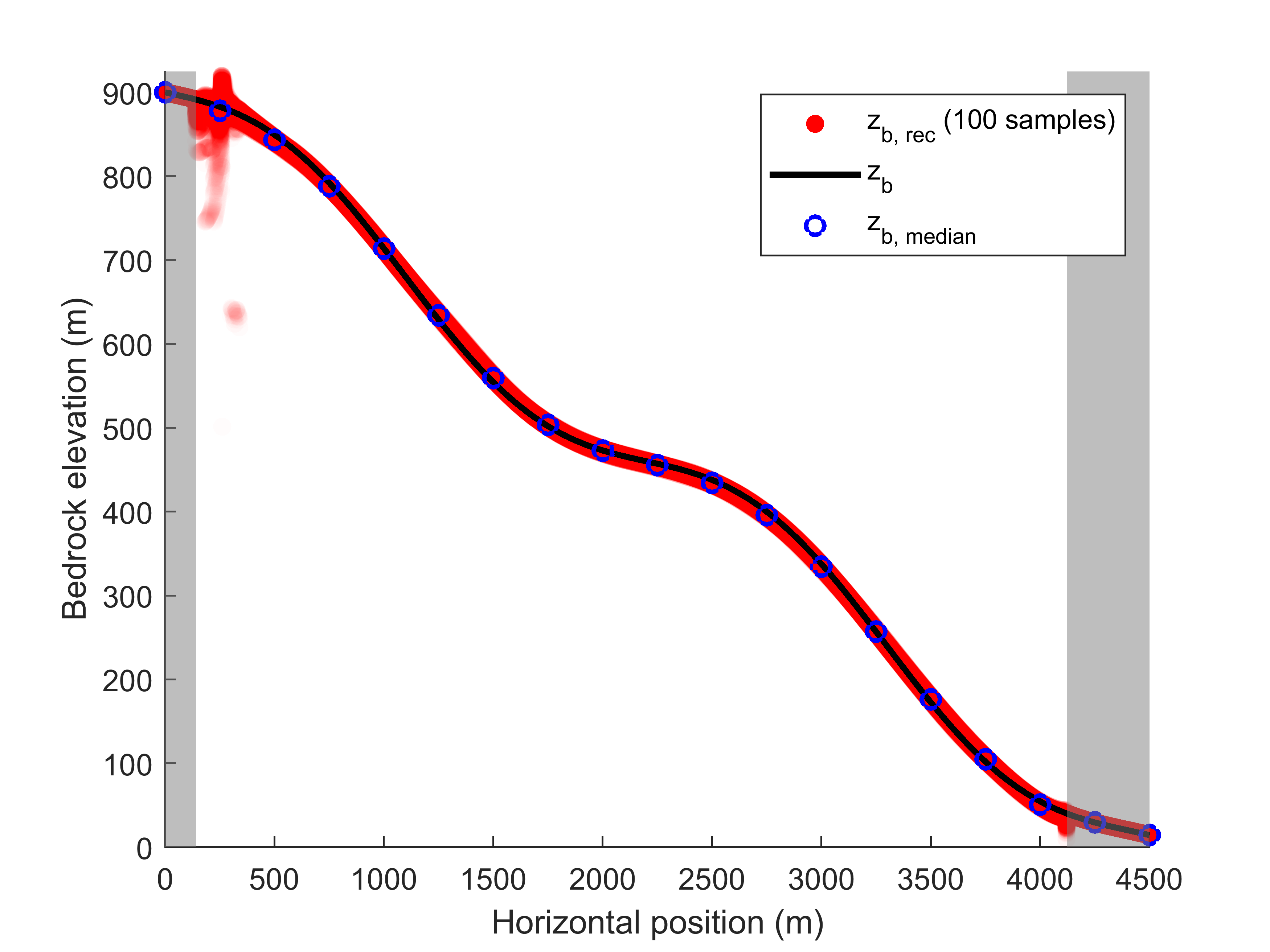}
    \subcaption[(a)]{}
\end{minipage}
\begin{minipage}{0.49\textwidth}
    \centering
    \includegraphics[width=\textwidth]{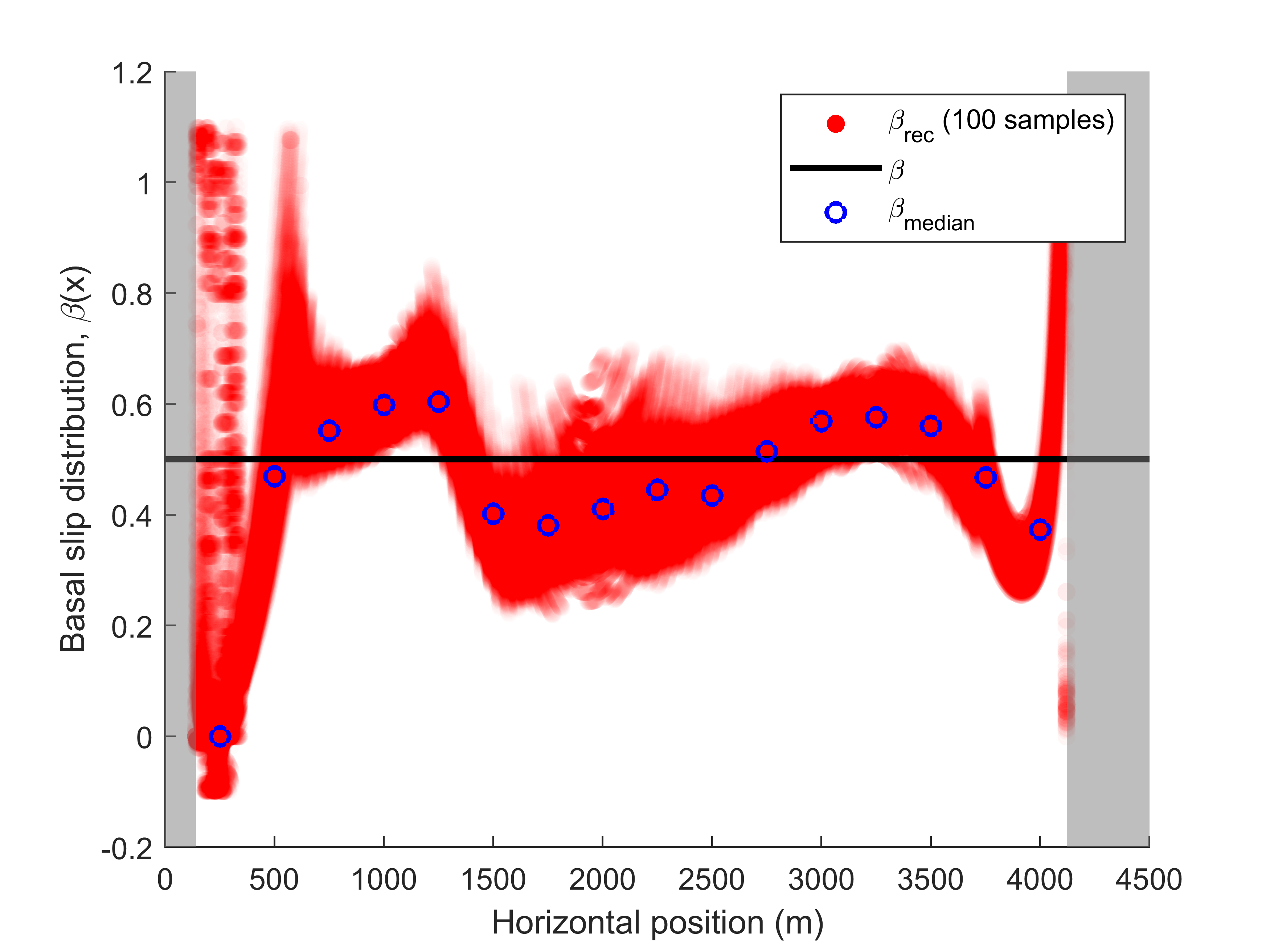}
    \subcaption[(b)]{}
\end{minipage}
\caption{Recovered bedrock elevation profile (a) and basal slip distribution (b) for 100 samples of noisy $S(x)$ where underlying bedrock is $z_b = b$ and $\beta = 0.5$. The solution envelope is given in red, the median solution as blue circles and the true as a black line for reference. In this case, $\epsilon = 0.1$.}
\label{fig:noise_set1_S}
\end{figure*}

\begin{figure*}[ht]
	\centering
	\begin{minipage}{0.49\textwidth}
		\centering
		\includegraphics[width=\textwidth]{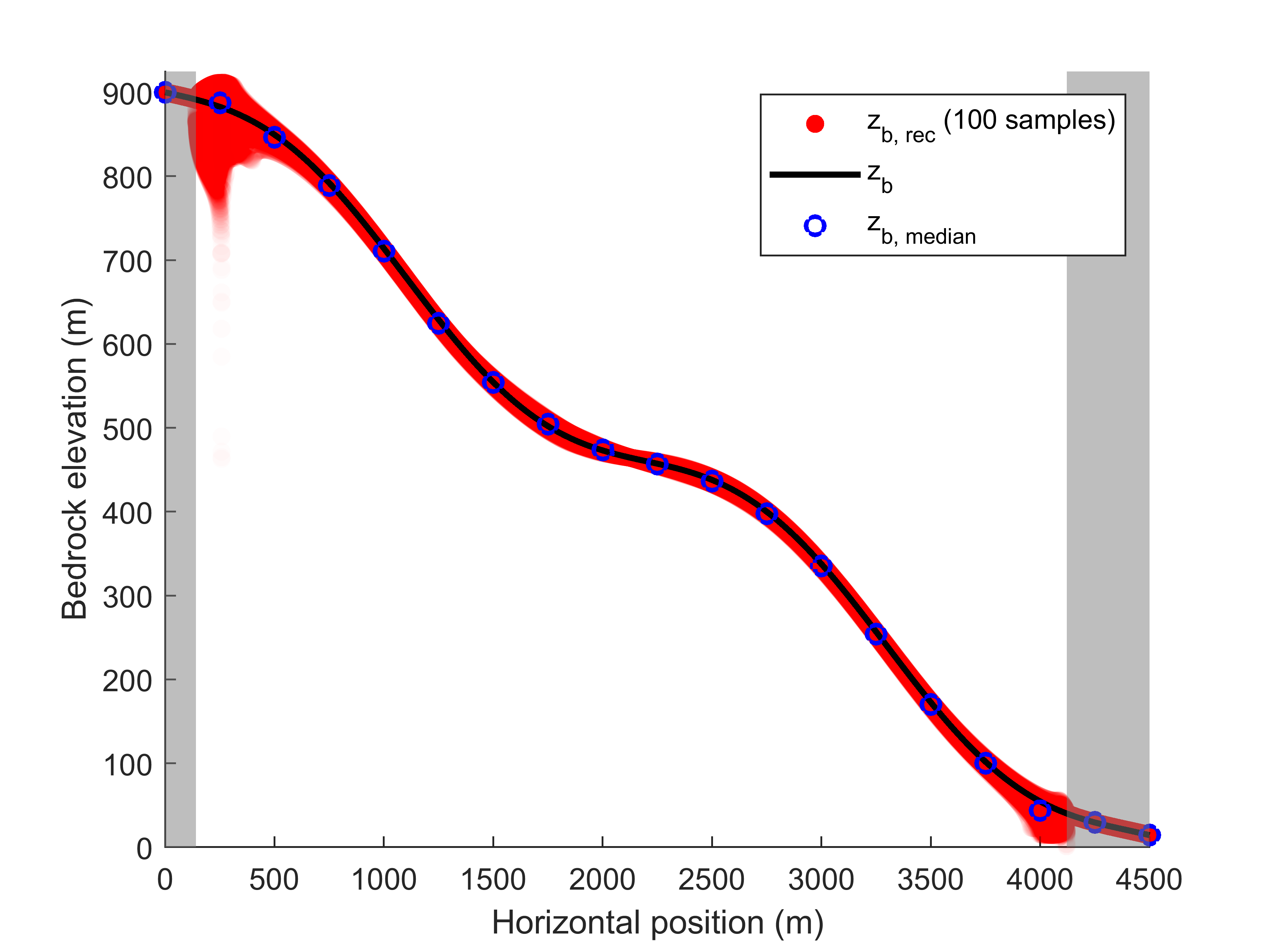}
		\subcaption[(a)]{}
	\end{minipage}
	\begin{minipage}{0.49\textwidth}
		\centering
		\includegraphics[width=\textwidth]{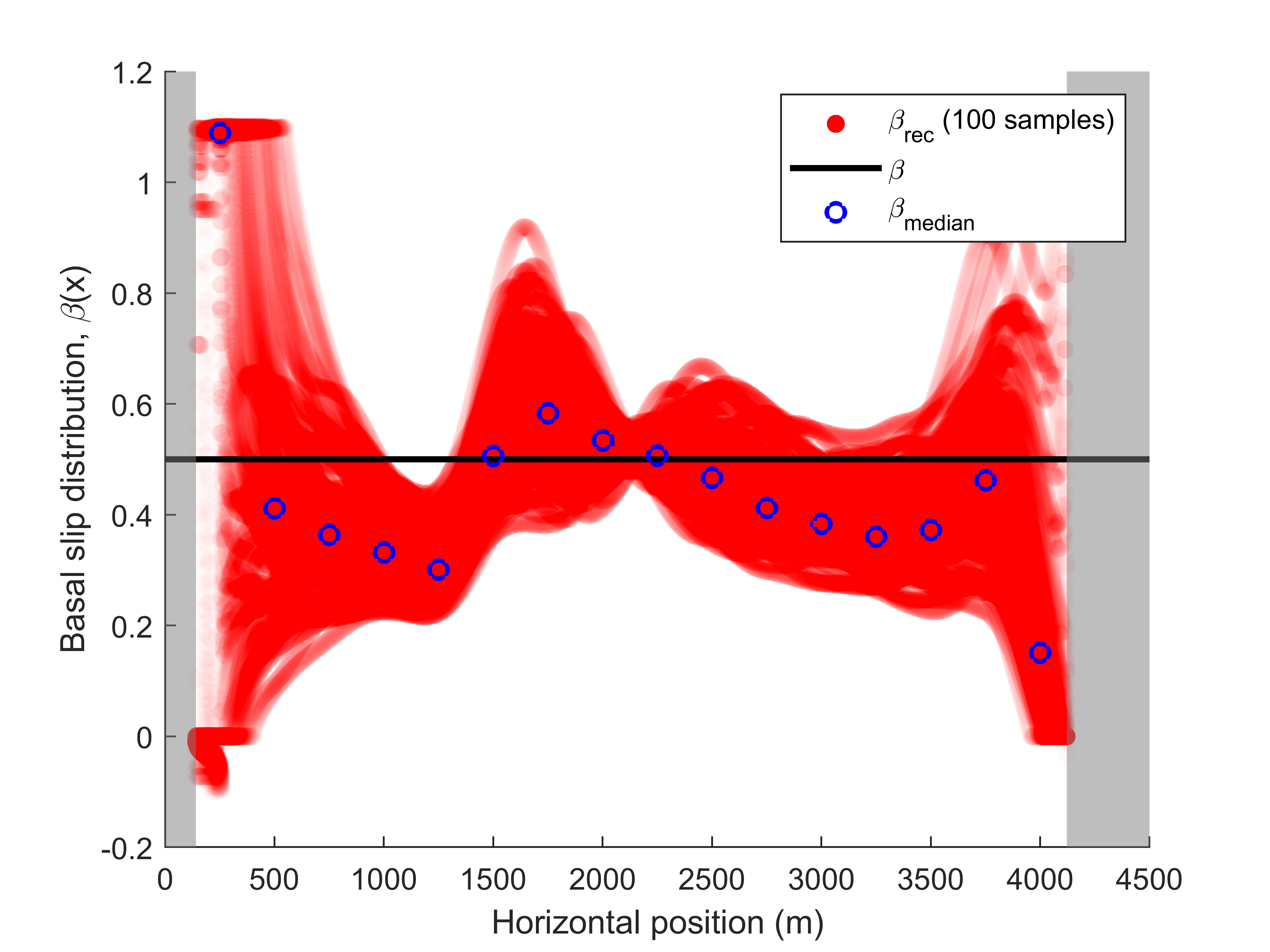}
		\subcaption[(b)]{}
	\end{minipage}
	\caption{Recovered bedrock elevation profile (a) and basal slip distribution (b) for 100 samples of noisy $u_s(x)$ where underlying bedrock is $z_b = b$ and $\beta = 0.5$. The solution envelope is given in red, the median solution as blue circles and the true as a black line for reference. In this case, $\epsilon = 0.1$.}
	\label{fig:noise_set1_us}
\end{figure*}

\begin{figure*}[ht]
	\centering
	\begin{minipage}{0.49\textwidth}
		\centering
		\includegraphics[width=\textwidth]{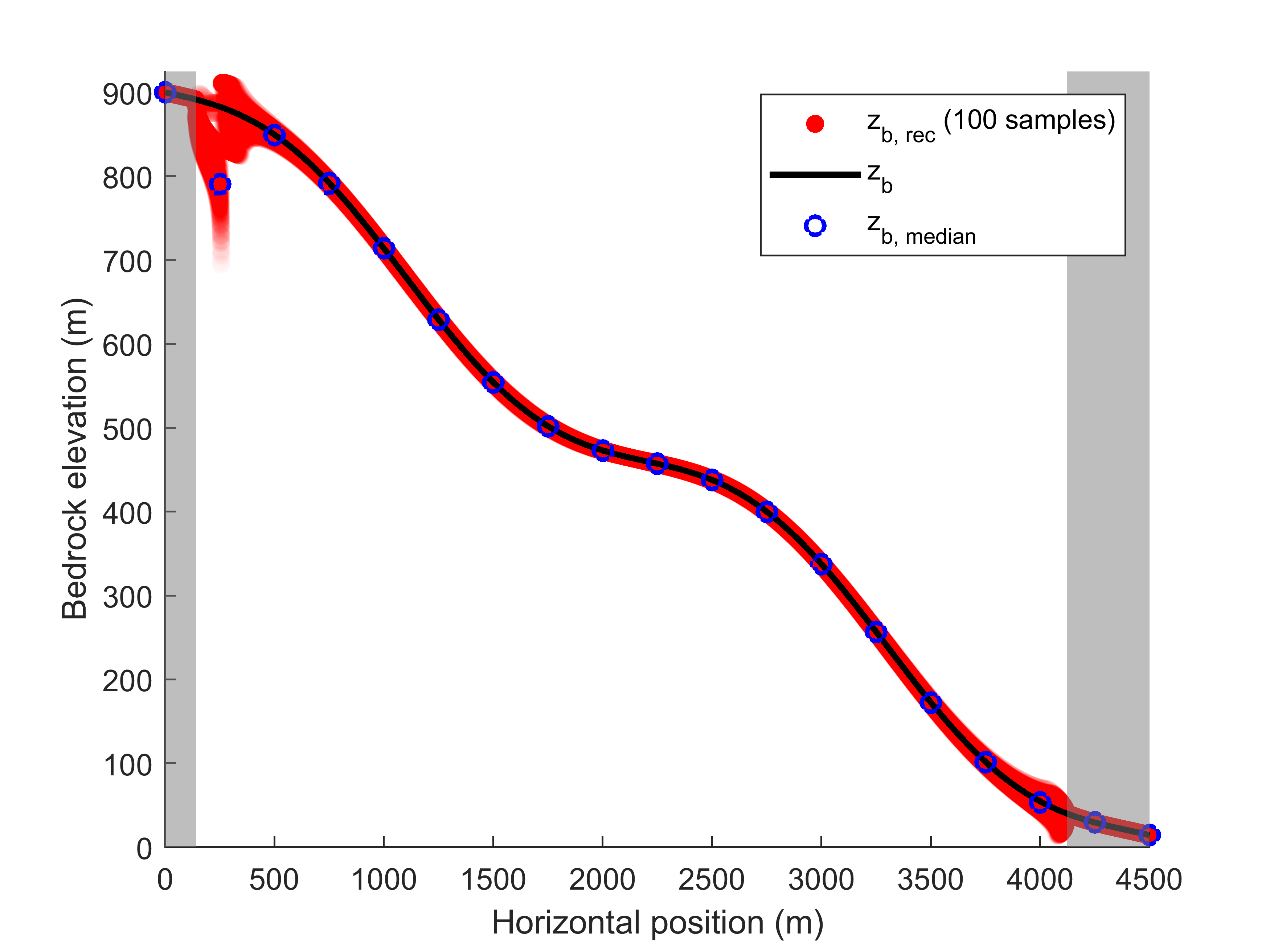}
		\subcaption[(a)]{}
	\end{minipage}
	\begin{minipage}{0.49\textwidth}
		\centering
		\includegraphics[width=\textwidth]{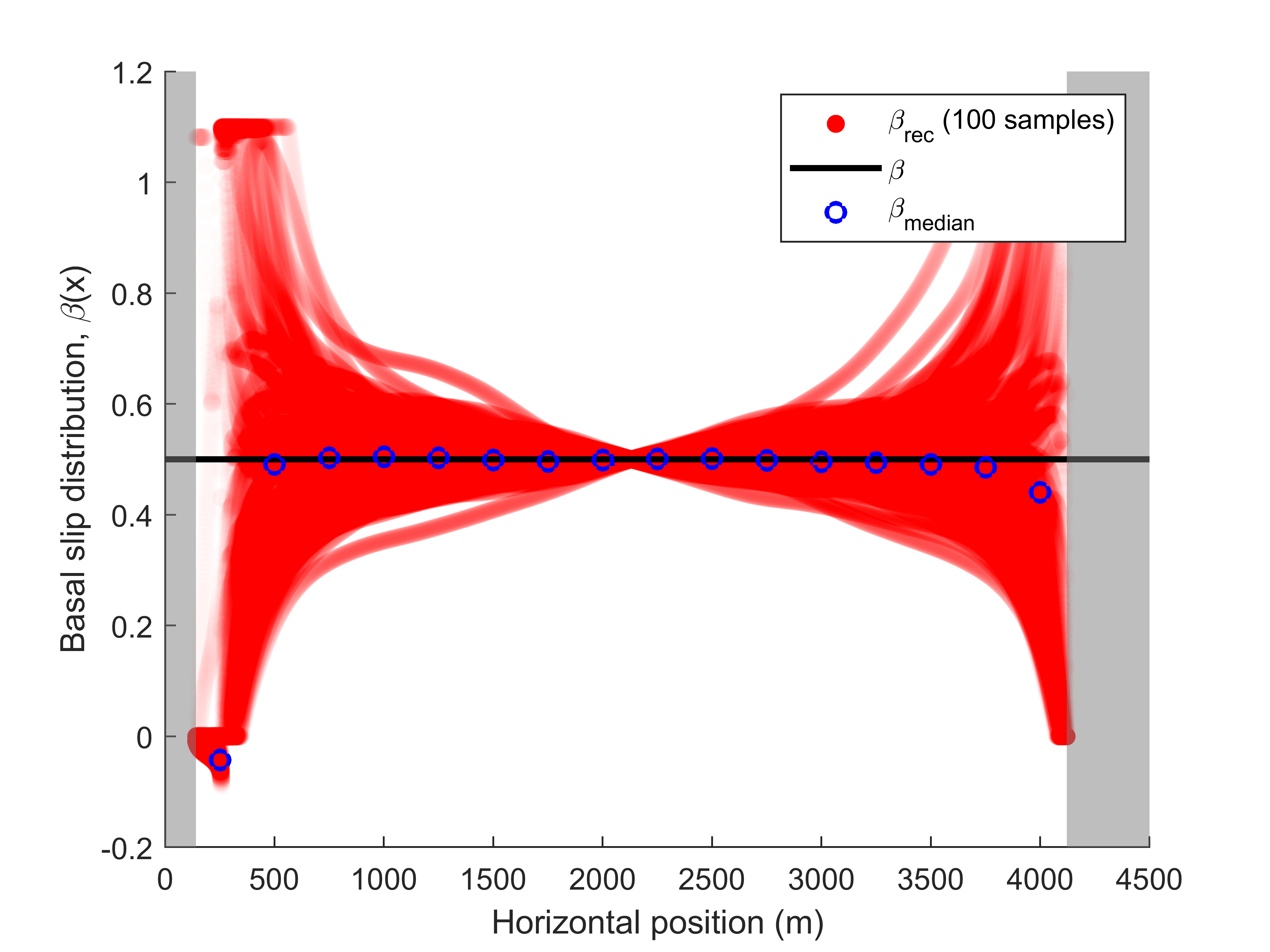}
		\subcaption[(b)]{}
	\end{minipage}
	\caption{Recovered bedrock elevation profile (a) and basal slip distribution (b) for 100 samples of noisy $a(x)$ where underlying bedrock is $z_b = b$ and $\beta = 0.5$. The solution envelope is given in red, the median solution as blue circles and the true as a black line for reference. In this case, $\epsilon = 0.1$.}
	\label{fig:noise_set1_a}
\end{figure*}

This process is applied to all benchmark cases with both $\epsilon = 0.1$ and $\epsilon = 0.2$. Examples of the results for 100 samples of noisy surface data using constant slip and a bumpy bedrock with $\epsilon = 0.1$ and $a_{offset} = 0$ are shown in Figs. \ref{fig:noise_set1_S} to \ref{fig:noise_set1_a}. Each figure is for inversion with a single noisy variable. Results for $\epsilon = 0.2$ and $a_{offset} = 0.1$ scale well so are not shown. The plots are created using a scatter function where each marker has a transparent fill. This helps to show where most solutions converged and which solutions were outliers based on depth of colour. The greyed areas show the outside the glacier domain where no ice is present. 

For noisy $S(x)$ and noisy $u_s(x)$, Fig. \ref{fig:noise_set1_S} and Fig. \ref{fig:noise_set1_us} respectively, the solution envelope for bedrock profile wavers around the true value, but does not become unstable, except for at the ice margins, see part Subfig. (a) in both cases. The envelope for the recovered basal slip distributions is much larger than the envelope for the bedrock profile for both. This is expected since recovered basal slip is very sensitive to errors in bedrock topography \citep{Gudmundsson2008}. Errors in height recovery, and hence bedrock topography, occur during the root solve in the polynomial (Eq. \eqref{eq:f(h)}) due to the terms involving $S(x)$ and $u_s(x)$, namely $C_0$, $\frac{\partial S}{\partial x}$, and $u_sH$. For noisy $S(x)$ wee see that the shape of the solution for basal slip in Fig. \ref{fig:noise_set1_S} (b) follows the size gradient, $\frac{\partial S}{\partial x}$. This is due to the noisy gradient multiplying powers of $H$ in the polynomial resulting in errors in bedrock following this shape which is then exaggerated when solving for basal slip. For noisy $u_s(x)$, the shape of the error we see in Fig. \ref{fig:noise_set1_us} (b) is less closely related to $\frac{\partial S}{\partial x}$ but that it must still has some impact due to the balancing that occurs to solve Eq. \eqref{eq:f(h)}.

\begin{figure*}[ht]
	\centering
	\begin{minipage}{0.49\textwidth}
		\centering
		\includegraphics[width=\textwidth]{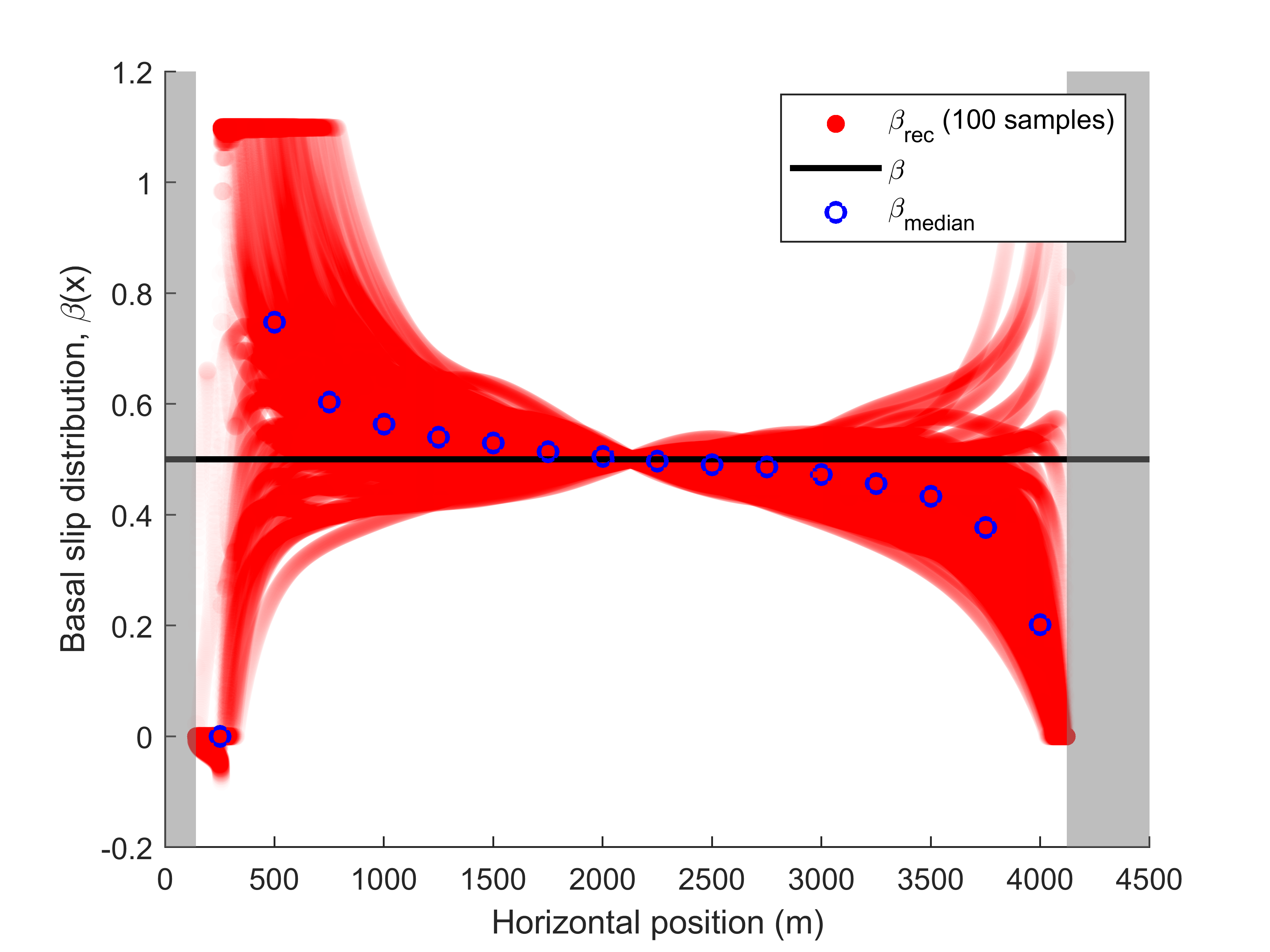}
		\subcaption[(a)]{}
	\end{minipage}
	\begin{minipage}{0.49\textwidth}
		\centering
		\includegraphics[width=\textwidth]{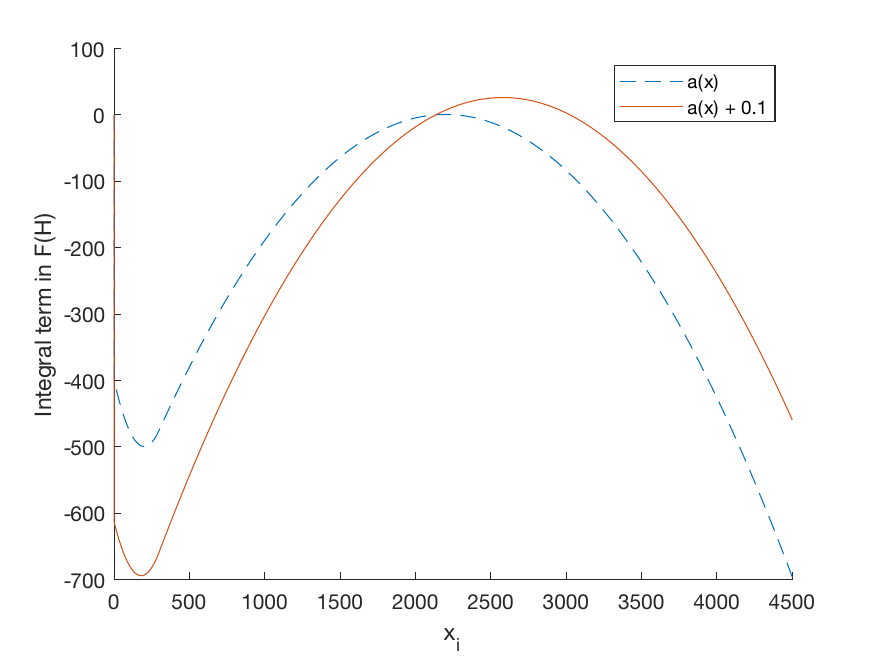}
		\subcaption[(b)]{}
	\end{minipage}
	\caption{Recovered basal slip distribution (a) for 100 samples of noisy $a(x)$ where underlying bedrock is $z_b = b$ and $\beta = 0.5$. The solution envelope is given in red, the median solution as blue circles and the true as a black line for reference. In this case, $\epsilon = 0.1$ and $a_{offset} = 0.1$. }
	\label{fig:noise_a_expl}
\end{figure*}

For noisy $a(x)$, the solution envelope for recovered bedrock elevation performs similarly to the prior two noisy variables. For basal slip however, the solution envelope exhibits some interesting behaviour, with large errors at the glacier margins but minimal error in the middle. Considering again our polynomial $F(H_i)$ in Eq. \eqref{eq:f(h)}, and expanding the $C_0$ term we get,
\begin{align}
F(H_i) = \int_{0}^{x_i} a dx - \int_{0}^{x_0} + \quad \text{terms with no noise}
\end{align}
which will explain the behaviour we see well. At $x_0$, the integral terms cancel which removes the effect of noise from the inverse method. This is clearly shown in Fig. \ref{fig:noise_set1_a} (b) where at $x_0$ we see that all noisy solutions have no error. To the left of $x_0$, the integral term simplifies to,
\begin{align}
\int_{0}^{x_i} a dx - \int_{0}^{x_0} a dx = -\int_{x_i}^{x_0} a dx
\end{align}
and to the right,
\begin{align}
\int_{0}^{x_i} a dx - \int_{0}^{x_0} a dx = \int_{x_0}^{x_i} a dx.
\end{align}
Hence we see behaviour in the solution which is responding to whether the integral term involving the noisy $a(x)$ is more or less than the true value. When the integral is more than expected, $H$ follows suit resulting in a smaller $\beta$ and similarly when the integral term is less than expected $\beta$ is smaller. This feature is clearly exhibited in noisy test case where $a$ is offset by 0.1. which is shown in Fig.\ref{fig:noise_a_expl} (a). The related integral term in $F(H)$ is plotted in alongside in Fig. \ref{fig:noise_a_expl} (b). While there is still some uncertainty in the basal slip solution, the median solution clearly exhibits the behaviour we expect.

\section{Discussion}
\label{sec:discussion}
For each combination of bedrock elevation profile and basal slip distribution, the recovery of each input variable was good. Errors in the reconstructed bedrock elevation, for all scenarios, was negligibly small. Similarly for the basal slip distribution, though there were some relatively larger errors in this recovery. The largest errors in all cases arose at either the top or bottom end of the glacier. 

At the top end of the glacier, there is a dome where the gradient of the free surface elevation is 0, in other words $\frac{\partial S}{\partial x} = 0$. Clearly, given Eq. (\ref{eq:us}) for surface velocity, this results in a stagnation point in the free surface. Due to this, no information about $\beta(x)$ or $H(x)$ can be recovered from Eq. (\ref{eq:us}). In Newton's method, this stagnation point presents as $F'(H) = 0$, which means that the method cannot proceed. Additionally, when $F'(H)$ becomes very small, we have an ill-conditioned problem and Newton's method may not converge. To combat this, when $|\frac{\partial S}{\partial x}| \approx 0$, the previous solution for $H$ is taken and the method skips to the next point.

At the bottom end of the glacier, $|\frac{\partial S}{\partial x}| \to \infty$. Because a finite number of points is used to approximate the derivative in places where it changes rapidly, such as at the bottom end, the approximation is worse. These approximation errors transfer across to the inverse solution. In addition, at the very top and very bottom where the ice ends, $\frac{\partial S}{\partial x}$ is discontinuous. This discontinuity gives rise to error also.

In the sensitivity analysis, it is clear that recovery of height is robust against noise whereas basal slip is far more sensitive to inputs. This aligns with \citet{Farinotti2017} which found that models relying on multiple inputs, as our coupled inverse method does, have high sensitivity to input data quality. It is promising however that the errors do have associated explanations which may be used to account for uncertainties in a next generation model. Hence, a proper pre-processing method is necessary for the well-conditioning of the problem.

Overall the method has performed well in the restricted, idealised cases tested here. The main caveats in considering the applicability to real cases are steady state assumption, the restriction to the SIA model, and the wavelength of $\beta$ considered. Further consideration should certainly be given to the steady state assumption as this will not apply to many cases in today's climate. An in-depth analysis using the method briefly explained in Subsec. \ref{subsec:non-steady-state} would be a worthwhile next step. Secondly, the SIA model is restrictive to slow moving grounded ice which restricts the uses for this method. It would be interesting to see if a similar techniques could be applied for ice stream/sheet models which experience large basal slip. Pairing the two may allow for further research into grounding line movement in Antarctic ice sheets. 

Finally, results in this paper are only given for test cases where the basal slip distribution has variation over large wavelengths. In preliminary modelling, distributions with shorter wavelengths were explored briefly. However, as \citep{Gudmundsson2008} found, small amplitude perturbations in basal slipperiness could only be detected in the surface measurements if the perturbation had a large wavelength in comparison with the ice thickness. If the wavelength was too small, mixing occurred in the surface data between basal slip and bedrock topography which caused the inverse method to fail in basal slip recovery. This restricts the ability of the inverse method to detect small wavelength changes in basal slip which are physically realistic for many ice flows.

\section{Conclusion}
\label{sec:conclusion}

The results show that it is possible to accurately recover both the bedrock elevation profile and basal slip distribution of a glacier for given surface elevation and velocity in certain realistic synthetic cases.
The simple method is robust regardless of the underlying bedrock profile and basal slip distribution.
This is a key result when considering the applicability of the method to `real world' problems in which bedrock and basal slip are unlikely to be uniform.
A logical next step in developing the method is testing performance in a three-dimensional flow.

Many previous authors have focused on bedrock recovery in no-slip cases.
That simplification gives rise to many interesting methods but neglects basal slip which plays a vital role in glacial evolution.
Indeed, since basal slip can drastically change the height of the glacier due to dynamic thinning, the bedrock elevation recovery can have large error if this is not considered.
In cases where basal slip is included, models to date have been complex. In contrast, the simple method presented here still returns the correct bedrock elevation profile and basal slip distribution with the more general inclusion of basal slip for certain broadly realistic synthetic cases. Further, a previously unknown implication is that a unique combination of bedrock elevation profile and basal slip distribution gives rise to unique surface elevation and velocity.  Further studies into the generalised case are required to prove this. 

\section*{Acknowledgements}

We would like to thank the reviewer for their thoughtful comments and efforts towards improving our manuscript.

\clearpage
\newpage



\bibliographystyle{spbasic}      
{\footnotesize\bibliography{acta_paper.bib}}   

\end{document}